\documentclass[aip,apm,reprint,amsmath,amssymb]{revtex4-1}

\usepackage{graphicx}
\usepackage{dcolumn}
\usepackage{bm}

\DeclareGraphicsExtensions{.pdf,.eps,.png,.jpg,.mps} 

\begin{document}

\preprint{}

\title{Spin-Polarized Electrons in Bilayer Graphene Flakes }

\author{P.A. Orellana}
\email{orellana@ucn.cl}
\affiliation{ Departamento de F\'{\i }sica, Universidad Cat\'{o}lica
del Norte, Casilla 1280, Antofagasta, Chile}

\author{L. Rosales}
\affiliation{Departamento de F\'{i}sica, Universidad T\'{e}cnica
Federico Santa Mar\'{i}a, Casilla 110-V, Valpara\'{i}so, Chile}

\author{L. Chico}
\affiliation{Departamento de Teor\'{\i}a y Simulaci\'on de Materiales, Instituto de Ciencia de Materiales de Madrid, CSIC, 28049 Cantoblanco, Spain}

\author{M. Pacheco}
\affiliation{Departamento de F\'{i}sica, Universidad T\'{e}cnica
Federico Santa Mar\'{i}a, Casilla 110-V, Valpara\'{i}so, Chile}

\date{\today}

\begin{abstract}
We show that a bilayer graphene flake deposited above a ferromagnetic insulator
can behave as a spin-filtering device.
The ferromagnetic material induces exchange splitting in the graphene flake, and 
due to the Fano antiresonances occurring in the transmission of the graphene flake as a function of flake length and energy, it is possible to obtain a net spin current.
This happens when an antiresonance for one spin channel coincides with a maximum transmission for the opposite spin. 
We propose these structures as a means to obtain spin-polarized currents and spin filters in graphene-based systems.

\end{abstract}



\maketitle

\section{Introduction}

In the last years, there has been much interest in 
exploring the 
unique properties of nanostructures 
for  spintronic devices, which utilize the spin degree of freedom of the electron as the basis of their operation \cite{Datta_1990}. 
To this end, novel ways of 
 generating and detecting spin-polarized currents have been explored.  
For instance, Song {\it et al.} \cite{Song_2003} described how a spin filter might be achieved in open systems by exploiting the Fano resonances occurring in their transmission characteristics. 
The Fano effect \cite{Fano_1961, Fano2} arises when quantum interference takes place between two competing pathways, one connecting with a continuum of energy states and the other with discrete states. In open systems in which the spin degeneracy of carriers has been lifted, 
this effect might be used as an effective means to generate spin polarization of transmitted carriers. The idea is to tune the system so that a transmission resonance for one spin channel coincides with an antiresonance for the opposite spin. In this way, a spin polarized current arises.

Previous works have shown that  graphene bilayer flakes exhibit Fano antiresonances in the transmission \cite{Jhon_2010, Jhon_2011}.  
We propose to exploit these antiresonances to produce spin polarized currents in a graphene-based system 
 by putting the graphene flake in contact with a magnetic insulator, such as EuO \cite{Haugen_2008,Munarriz_2012}.  Exchange splitting is induced in the graphene flake due to 
the magnetic proximity effect \cite{Tedrow_1986,Tkaczyk_1988,Hao_1991}.
This produces an opposite energy shift in the spin-up and down antiresonances in the conductance, yielding a spin polarization of the current. 
The feasibility of spin-polarized currents in graphene devices is important for the development of all-graphene electronics, which is one of the goals in the research focused on graphene applications \cite{Hicks_2012}. Recently, it has been experimentally demonstrated the growth of  Eu on graphene, thus bringing closer the possibility of the  proposed scheme to induce spin polarized currents in graphene-based systems. \cite{Forster_2012}
   
\begin{figure}
\includegraphics*[angle=-90,width=\columnwidth,clip]{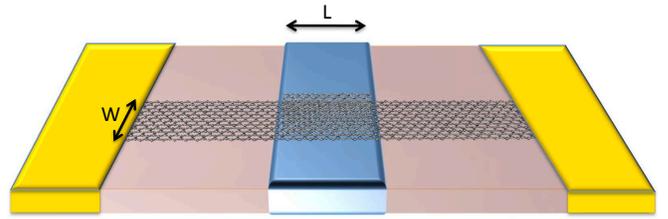}
\caption{Schematic view of a graphene flake deposited onto a graphene nanoribbon. This system can be considered as a bilayer flake with monolayer ribbon contacts. 
The bilayer portion of the system is placed over a ferromagnetic material (blue).
 } \label{fig:flake}
\end{figure}

In this work we study the transport properties of a ferromagnetic bilayer graphene flake, showing that it can behave as a spin filter.  We consider that the flake is contacted by monolayer nanoribbons, and the contacts are connected to the same layer of the flake, as shown in figure \ref{fig:flake}.  Such configuration can be achieved by placing a monolayer flake onto a nanoribbon. It is also possible to form a bilayer flake by overlapping two semi-infinite nanoribbons \cite{Jhon_2010,acta_2012}; however, we concentrate here in the bottom-bottom geometry, given that the changes due to the configuration of the contacts merely produce a variation in the position of the transmission antiresonances \cite{Jhon_2010, Jhon_2011}. We employ a one-orbital tight-binding model which we solve analytically within the single-mode approximation. This solution agrees perfectly with the numerical result obtained by a Green function method in the one-mode energy range.

Our main results are the following: \\
(i) We obtain an analytical solution for the spin-dependent transmission through bilayer graphene flakes. The comparison of this analytical result to the numerically computed transmission, obtained by a recursive Green function method, is excellent in the one-mode energy range. \\
(ii) The analytical expressions for the transmission allows us to explore thoroughly the parameter space, locating the most advantageous system sizes to obtain a net spin current. \\
(iii) We have found that the maximum spin polarization is obtained when sharp antiresonances are produced in a plateau with maximum transmission. These correspond to quasi-localized states in the bilayer graphene flake. Thus, the tuning of the flake length is important to obtain a net spin current.

\begin{figure}
\centerline{\includegraphics[width=90mm,clip]{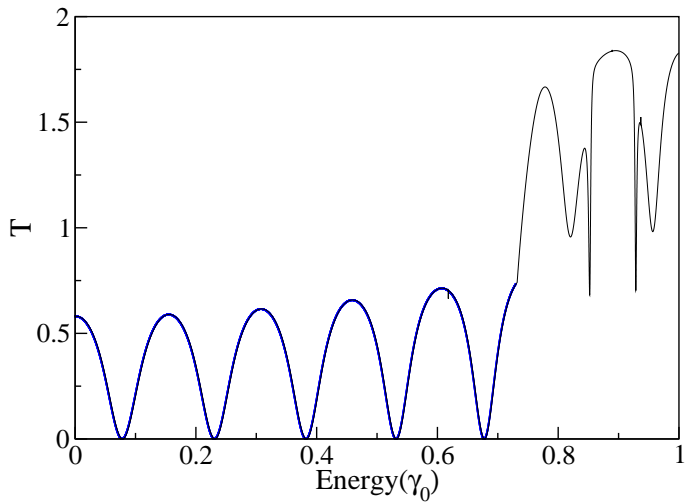}}
\caption{Transmission versus energy for spin-up carriers of the armchair bilayer flake of width $W=5$ and length $L=30$ (black thin line) and the coupled-chain model of equivalent length,  $n=120$ (blue thick line). The latter is shown only for the one-mode energy interval of the armchair ribbon system.}
\label{fig:compar}
\end{figure}

\section{Model}

As we focus on the electronic properties close to the Fermi energy, we employ a $\pi$-orbital tight-binding model, which gives an excellent description of the low-energy properties of graphene systems.  We consider the in-plane nearest-neighbor interaction given by a single hopping parameter $\gamma_0$, which is approximately 3 eV, concentrating on metallic armchair nanoribbons, which can play the role of contacts in this system. 
We would like to note that, in principle, all armchair nanoribbons show a gap due to edge reconstruction, as evidenced by first-principles calculations \cite{Son}. A change of the hopping parameter at the edges can reproduce this effect within the one-orbital approximation \cite{Son,Gunlycke}; however, we choose all hoppings to be equal in order to solve analytically the problem and then compare to the numerical calculations. Just above the gap, which diminishes with the ribbon width, the behaviour is as reported here.

With respect to the bilayer flake, we focus in one type of stacking, the AA, taking into account only one hopping parameter $\gamma_1$ between atoms placed directly on top of each other.  We assume that $\gamma_1=0.1 \gamma_0$. Note that, although the AB stacking is more stable for graphite, the direct or AA stacking has been experimentally found in few-layer graphene \cite{Liu_2009}. Moreover, the antiresonances which give rise to the energy windows where spin polarization takes place in these systems are present both in AA- or in AB-stacked bilayer graphene flakes, so the physics of the problem can be studied in any of the two geometries. Our choice of the AA stacking is motivated by that fact that, as we show in this paper, the AA-stacked system can be easily mapped into a one-dimensional chain with two hoppings, yielding analytical expressions of the transmission that perfectly fit the conductance obtained numerically {\it in the single mode regime}. Within this description, we measure the length of the flake in translational unit cells, which corresponds to the number of 4-atom units along the flake. For the width we use the standard notation, giving the number of dimer chains across the flake. 

 The Hamiltonian of the complete system can be written as a sum of the left and right lead parts plus a central part, which constitutes the bilayer flake:

\begin{equation}
\mathcal{H}= \mathcal{H}_L+  \mathcal{H}_R + \mathcal{H}_F
\end{equation}
with 
\begin{equation}
\mathcal{H}_{L,R}= 
- \gamma_0\sum_{\langle i,j\rangle , \sigma} c_{i,\sigma} ^{\dagger}c_ {j,\sigma}, 
\end{equation}
where  $c_{i,\sigma} ^{\dagger}$ ($c_{i,\sigma}$) stands for the creation (annihilation) operator of an electron with spin $\sigma$ on site $i$, and the sum $\langle i,j\rangle $ is over nearest neighbors.
The Hamiltonian of the  central conductor, i.e., the bilayer graphene flake, is given by

\begin{eqnarray}
\mathcal{H}_{F}= 
- \gamma_0\sum_{\langle i,j\rangle , \sigma,\alpha} c_{i,\sigma,\alpha} ^{\dagger}c_ {j,\sigma,\alpha}
- \gamma_1\sum_{ i, \sigma} (c_{i,\sigma,u} ^{\dagger}c_ {i,\sigma,d} + h.c.)\nonumber\\
+\sum_{i, \sigma, \alpha }   \sigma\,\Delta_\mathrm{ex}c^{\dagger}_{i,\sigma,\alpha}c_{i,\sigma,\alpha} \,\,\,\,\,\,\,\,\,\,\,
\end{eqnarray}

\noindent
In this latter expression $\alpha$ runs over $u$, $d$, i.e., the upper and lower flake, respectively, the spin index $\sigma$ takes the values $\pm 1$, and $\Delta_\mathrm{ex}$ is the exchange splitting induced by the ferromagnetic insulator. For graphene systems on EuO we take the value $\Delta_\mathrm{ex}=0.001\gamma_0$ \cite{Haugen_2008,Munarriz_2012}. For the sake of simplicity in the analytical approach, we assume that the 
same ferromagnetic term is included in both layers of the flake. However, we have numerically computed both possibilities (i.e., spin interaction in one or in two layers) and found  the main difference is that the energy shift of the antiresonances is twice when the ferromagnetic coupling affects the two layers.

The transport properties of this structure can be obtained numerically by computing the Green function of the system, which yields the conductance in the Landauer-Kubo approximation \cite{Datta_book}. 
However, as this system is uniform with a high symmetry, one can take advantage of this and employ an analytical approach to solve the problem in the single-mode approximation. 

The tight-binding equations of motion for the amplitude of probability to find an electron in site $i(j)$ of atom $A(B)$ in a bilayer ribbon flake are given by
\begin{widetext}
\begin{eqnarray}
(E - \sigma\,\Delta_\mathrm{ex})\psi^{A,\alpha}_{j,m,\sigma}&=&\gamma_0(\psi^{B,\alpha}_{j,m,\sigma}+\psi^{B,\alpha}_{j-1,m-1,\sigma}+\psi^{B,\alpha}_{j-1,m+1,\sigma}) +\gamma_1 \psi^{A,\bar{\alpha}}_{j,m,\sigma}\nonumber \\
(E- \sigma\,\Delta_\mathrm{ex})\psi^{B,\alpha}_{j,m,\sigma}&=&\gamma_0(\psi^{A,\alpha}_{j,m,\sigma}+\psi^{A,\alpha}_{j+1,m-1,\sigma}+\psi^{A,\alpha}_{j+1,m+1,\sigma})+\gamma_1 \psi^{B,\bar{\alpha}}_{j,m,\sigma}.
\label{eq:motion}
\end{eqnarray}
\end{widetext}
Assuming that the solutions in the $y$-direction are a combination of plane waves of the form $e^{\pm i m q}$, 
the expressions (\ref{eq:motion}) are reduced to the equations of motion of two coupled dimer chains (see Appendix), which eventually lead to a spin-dependent transmission given by
\begin{widetext}
\begin{equation}
T_\sigma(E)=\frac{(4-\frac{E^2}{\gamma_0^2})G_\sigma^{2}} {[(G_\sigma^{2}-F_\sigma^{2}-1)\frac{E}{2\gamma_0}-2F_\sigma]^2+(G_\sigma-F_\sigma+1)^{2}(1-\frac{E^2}{4\gamma_0^2})},
\label{trans}
\end{equation}
\end{widetext}
where $G_\sigma$, $F_\sigma$ are given by 
\begin{eqnarray}
G_\sigma & = &  \frac{1}{2} \left[ U_n^{-1} \left(\frac{E_{\sigma,+}}{2}\right) + U_n^{-1}  \left(\frac{E_{\sigma,-}}{2}\right)  \right]  \nonumber \\
F_\sigma & =  &\frac{1}{2} \left[ \frac{U_{n-1} \left(\frac{E_{\sigma,+}}{2}\right)}{U_n  \left(\frac{E_{\sigma,+}}{2}\right)} + 
\frac{U_{n-1}\left(\frac{E_{\sigma,-}}{2}\right)}{U_n \left(\frac{E_{\sigma,-}}{2}\right)}
\right].
\end{eqnarray}
In the above $U_n(x)$ are the Chebyshev polynomials of the second kind and  $E_{\sigma, \pm} = E -  \sigma\,\Delta_\mathrm{ex} \mp \gamma_1$.

In order to quantify the spin-dependent transport, it is necessary to evaluate the degree of spin polarization of the electric current. There are several ways to do this, but following Ref.~\cite{Song_2003} we introduce the weighted spin polarization
as:
\begin{equation}
P_{\sigma }=\frac{\left\vert T_{\uparrow }-T_{\downarrow }\right\vert }{
\left\vert T_{\uparrow }+T_{\downarrow }\right\vert }\,T_{\sigma }.
\label{wsp}
\end{equation}
Notice that this definition takes into account not only the relative
fraction of one of the spins, but also the contribution of those spins to
the electric current. In other words, we will require that not only the
first term of the right-hand side of~(\ref{wsp}) to be of order of unity,
but also the transmission probability $T_{\sigma }(E)$ should be significant.

\begin{figure}
\centerline{\includegraphics[width=95mm,clip]{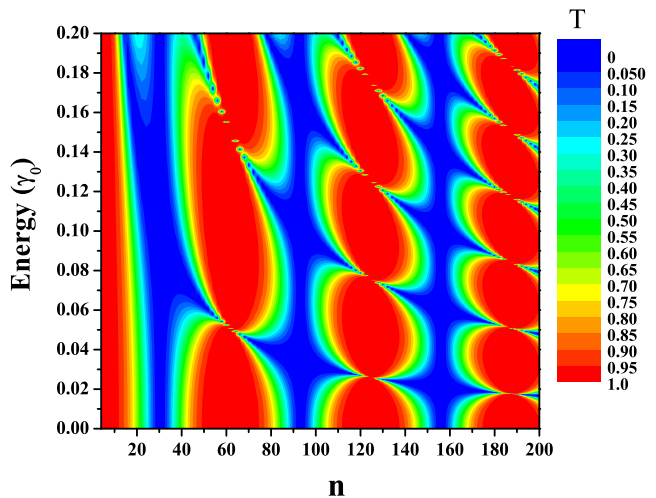}}
\caption{Contour plot of the transmission versus energy and coupled-chain length $n$ without ferromagnetic gate  ($\Delta_\mathrm{ex}=0$).}
\label{fig:transcp}
\end{figure}

\begin{figure}
\centerline{\includegraphics[width=90mm,clip]{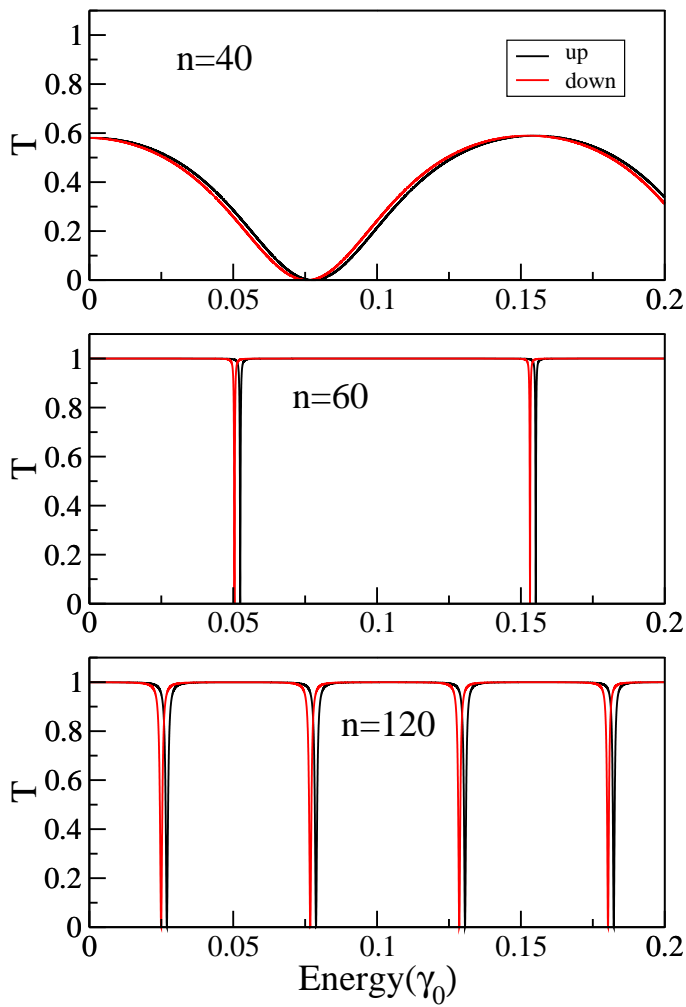}}
\caption{(Color online) Spin-dependent transmission for the graphene flakes of length $n=40,60$ and 120  with $ \Delta_\mathrm{ex}=0.001\gamma_0$.  The solid black line is for 
spin up, and the  solid red line represents  the spin down.
}
\label{fig:transspin}
\end{figure}

\section{Results}

We begin by comparing the analytical results for the one-dimensional chain to the conductance of the graphene nanoribbon system obtained numerically employing a decimation technique to obtain the Green function of the structure, as done previously for similar systems  without spin polarization \cite{Jhon_2010,Jhon_2012}. Given that the shift produced by the effective exchange is small, we just present the results for the spin-up channel. As the armchair ribbon length is measured in units comprised of four atoms, the equivalence between the one-dimensional chain and the armchair ribbon length is $n=4N$. 
Fig. \ref{fig:compar} shows the transmission for the $n=120$ one-dimensional chain model along with the transmission for the armchair nanoribbon system of width $W=5$ (two complete hexagons) and bilayer flake length $N=30$. The energy interval is chosen to display the results for the armchair system with two modes contributing to the total transmission.  As expected, the results for the one-dimensional chain transmission and the one obtained for the armchair nanoribbon system perfectly coincide
in the whole one-mode energy window. Thus, in the following, we focus on the analytic results in the single-mode approximation.

\begin{figure}
\centerline{\includegraphics[width=90mm,clip]{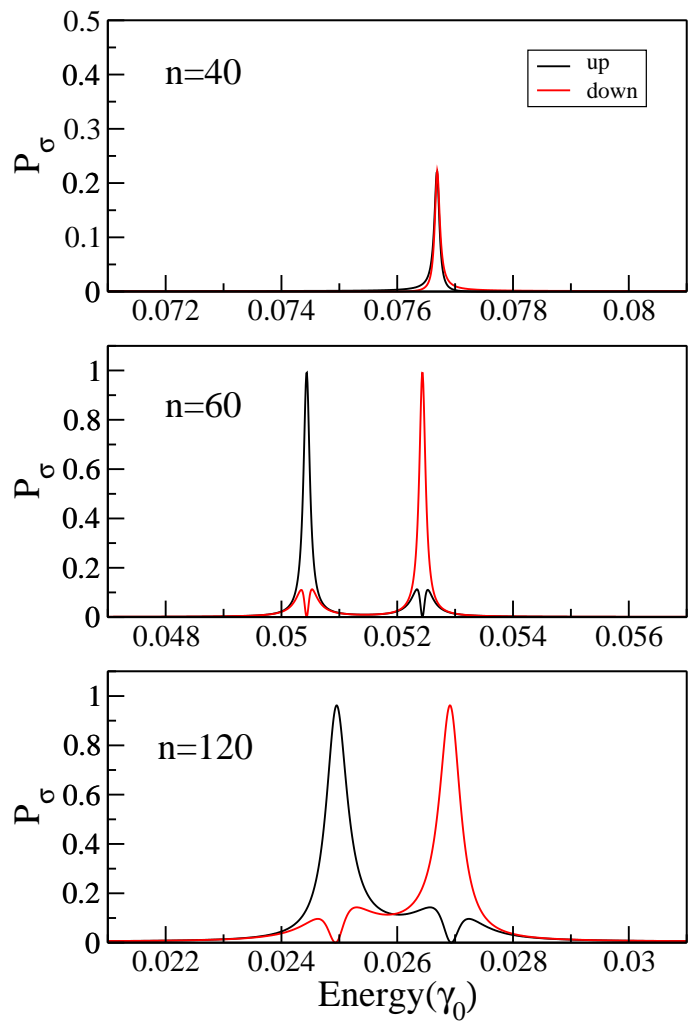}}
\caption{Spin polarization versus energy for different  flake lengths, with  $\Delta_\mathrm{ex}=0.001\gamma_0$. }
\label{fig:pol}
\end{figure}

\begin{figure}
\centerline{\includegraphics[width=95mm,clip]{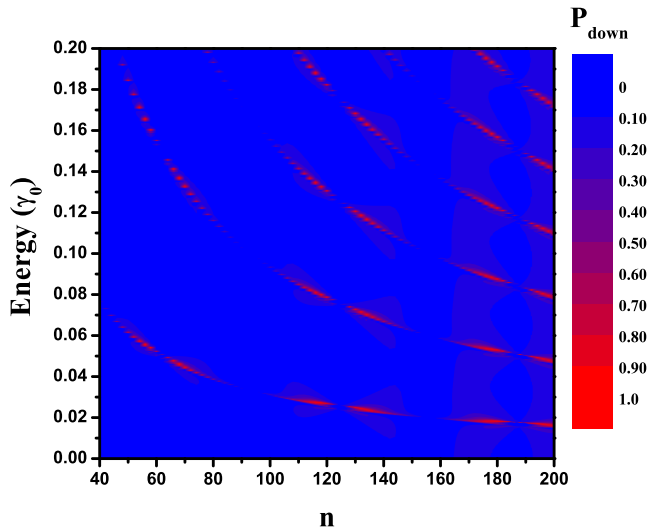}}
\caption{(Color online) Contour plot of weighted spin down polarization 
for the bilayer graphene flake with $\Delta_\mathrm{ex}=0.001\gamma_0$,  as a function of the Fermi energy and flake length.}
\label{fig:poldwcp}
\end{figure}

The analytical solution of the one-dimensional coupled chain allows us for exploring thoroughly the dependence of the transmission coefficient on the  flake length $n$ as a function of energy. 
As the exchange energy is very small, producing a shift of a few $meV$ in the antiresonances or any other features of the conductance, it is sufficient to study the transmission for zero exchange energy in order to identify the situations more convenient for spin filtering. 
Figure \ref{fig:transcp} presents a contour plot of the transmission for the one-dimensional coupled chains as a function of energy and flake size for $\Delta_\mathrm{ex}=0$. 
Within this single-mode approximation, our results present an excellent agreement to those obtained within the continuum Dirac model for graphene \cite{Jhon_2010}.

 From this figure \ref{fig:transcp} it is clear that for flake length $n= 30$ there is a minimum in the transmission, and that the sharpest antiresonances occur for $n= 60$, slightly above 120 and so on, where spin polarization will be maximized. 
 The $n=40$ has a noticeable oscillation of the transmission, but does not reach the maximum value of 1.
 To see more clearly this effect, we choose three lengths for the flake, namely, 40, 60, and 120,  and plot the transmission versus energy for the system placed over a ferromagnet, as described above. Figure \ref{fig:transspin} shows the transmission for spin-up and down electrons in these three cases. 
  We see that the $n=40$ case has a smoother oscillation that does not rise to transmission 1; 
 however, both $n=60$ and $n=120$ reach the maximum value of the conductance and have very sharp antiresonances, as anticipated from the contour plot.

For these three flake lengths, we compute the spin polarization as indicated in equation (\ref{wsp}). Even for $n=40$, where the transmission oscillates smoothly and the minimum values for opposite spins are barely distinguishable, there is a non-negligible value of the polarization, depicted in the top panel of figure \ref{fig:pol}. However, there is not appreciable spin filtering, given that the two peaks overlap in energy. For the other flake lengths, namely, $n=60$ and $n=120$,  the antiresonances in the transmission are very sharp and well-separated in energy. This yields a maximum polarization, shown in the center and bottom panels of figure  \ref{fig:pol}.

The analytical solution allows us to explore thoroughly the system sizes and energy ranges, with the aim to identify the optimal cases for spin filtering. Figure \ref{fig:poldwcp} is a contour plot of the weighted spin down polarization as a function of energy and bilayer flake length. As the spin-up and down antiresonances are very close in energy, this plot is sufficient to identify the cases with maximal weighted polarization. The maxima of the polarization contour plot follow lines that correspond to quasi-localized states that give rise to sharp antiresonances. 

It is enlightening to compare this contour plot to the one depicting the transmission (figure \ref{fig:transcp}), where we can observe vertical wide bands of maximum transmission (in red) that shrink for some values of energy. The lines drawn by the polarization maxima go precisely through the points where the maximal transmission regions narrow dramatically into a marked minimum. 

In the literature, it is has been showed that the Fano effect is robust under the several conditions, such as, the presence of impurities, electron-electron interaction, external fields, etc.\cite{Fano2}. In this sense, we believe that our model captures the essence of this phenomenon, which allow us to propose this spin filter device.

\section{Summary}

We have studied the transport properties of a bilayer graphene flake placed in close proximity to a ferromagnetic insulator. This produces an effective Zeeman splitting that can be exploited to obtain a spin filtering effect. 
We have shown that the tight-binding model for the ferromagnetic flake can be mapped into a one-dimensional chain that can be analytically solved in the single mode approximation. The analytical solution allows for a thorough exploration of the system parameters and the subsequent identification of the most convenient system sizes for a maximum spin polarization. 

\appendix
\section{Derivation of the transmission for the coupled linear chains}

In this Appendix we give in more detail the procedure to derive the spin-polarized transmission (\ref{trans}) analytically. 
We use the plane-wave form of the solutions 
\begin{eqnarray}
\psi^{A,\alpha}_{j,m,\sigma}&=&A^{q,\alpha}_{j,\sigma}e^{\pm i  m q}\nonumber \\
\psi^{B,\alpha}_{j,m,\sigma}&=&B^{q,\alpha}_{j,\sigma}e^{\pm i m q},
\label{planew}
\end{eqnarray}
which, substituted in the equations of motion (\ref{eq:motion}), yield
\begin{eqnarray}
(E-\epsilon_{j,\sigma})A^{q,\alpha}_{j,\sigma}&=&\gamma_0(B^{q,\alpha}_{j,\sigma}+2\cos q B^{q,\alpha}_{j-1,\sigma})+\gamma_1 A^{q,\bar{\alpha}}_{j,\sigma}\nonumber \\
(E-\epsilon_{j,\sigma})B^{q,\alpha}_{j,\sigma}&=&\gamma_0(A^{q,\alpha}_{j,\sigma}+2\cos q A^{q,\alpha}_{j+1,\sigma})+\gamma_{1} B^{q,\bar{\alpha}}_{j,\sigma}\;\;\;\;\;\;\;\;\;
\label{eqm2}
\end{eqnarray}

We restrict our calculation to the case of metallic armchair nanoribbons and lower energies, close to the Dirac point ($q=\pi/3)$. For this case, (\ref{eqm2}) can be mapped into the equations for two linear chains by replacing $A^{q,\alpha}_{j,\sigma}=f^{\alpha}_{l,\sigma}$, $B^{q,\alpha}_{j,\sigma}=f^{\alpha}_{l+1,\sigma}$,
\begin{eqnarray}
(E-\sigma \Delta_\mathrm{ex})f^{u}_{l,\sigma}&=&\gamma_0(f^{u}_{l+1,\sigma}+f^{u}_{l-1,\sigma})+\gamma_{1} f^{d}_{l,\sigma}\nonumber \\
(E-\sigma \Delta_\mathrm{ex})f^{d}_{l,\sigma}&=&\gamma_0(f^{d}_{l+1,\sigma}+f^{d}_{l-1,\sigma})+\gamma_{1} f^{u}_{l,\sigma}\;\;\;
\end{eqnarray}

Employing the symmetric and antisymmetric combination of the solutions,  $f^{\pm}=f^{u}\pm f^{d}$, the equations of motion become
\begin{eqnarray}
(E-\sigma \Delta_\mathrm{ex}-\gamma_{1})f^{+}_{l,\sigma}&=&\gamma_0(f^{+}_{l+1,\sigma}+f^{+}_{l-1,\sigma})
\nonumber \\
(E-\sigma \Delta_\mathrm{ex}+\gamma_{1})f^{-}_{l,\sigma}&=&\gamma_0(f^{-}_{l+1,\sigma}+f^{-}_{l-1,\sigma}).
\end{eqnarray}

In order to obtain the transmission through the system 
we consider open boundary conditions.  Plane waves incoming from
the left with amplitude unity are reflected from the scattering region (the bilayer flake) with amplitude $r_{\sigma}$, and are transmitted to the right contact with amplitude $t_{\sigma}$:
\begin{eqnarray}
f^{d}_{l,\sigma}=e^{ik l}+r_{\sigma} e^{-i k l},\,\, l<0\nonumber\\
f^{d}_{l,\sigma}=t_{\sigma} e^{i k l},\,\, l>N.
\end{eqnarray}

This leads to a system of equations on the reflection and transmission amplitudes. The transmission is given by $T_\sigma (E)=|t_\sigma|^2$. 

\begin{acknowledgments}
This work has been partially supported by the Spanish DGES under
grant FIS2009-08744, Chilean FONDECYT grants 1100560 (P.O.), 11090212 (L. R.) , and 1100672 (M.P.), and DGIP/USM internal grants 11.12.17 (L. R.) and 11.11.62 (M. P.). 
L. C. gratefully acknowledges the hospitality of the Universidad T\'ecnica Federico Santa Mar\'{\i}a (Chile). 
P.O., L. R. and M. P. acknowledge the warm hospitality of ICMM-CSIC (Spain) during July 2012. 
\end{acknowledgments}

\end{document}